\documentstyle[12pt,epsf,feynmf]{article}

\def\siml{{\ \lower-1.2pt\vbox{\hbox{\rlap{$<$}\lower6pt\vbox{\hbox{$\sim$}}}}\ }} 
\def\bfnabla{\mbox{\boldmath $\nabla$}}

\def\bfsigma{\mbox{\boldmath $\sigma$}}

\def\als{\alpha_{s}}
\def\al{\alpha}

\newcommand{\nn}{\nonumber}
\newcommand{\be}{\begin{equation}}
\newcommand{\ee}{\end{equation}}
\newcommand{\bea}{\begin{eqnarray}}
\newcommand{\eea}{\end{eqnarray}}

\def\dsl{\,\raise.15ex\hbox{/}\mkern-13.5mu D}

\newcommand{\MS}{\overline{\rm MS}}

\newcommand{\Appendix}[1]%
    {%
     \section{#1}%
      }

\begin{document}\setlength{\unitlength}{1mm}

\begin{fmffile}{dib3}

\begin{titlepage}
\begin{flushright}
\tt{UB-ECM-PF 04/37}
\end{flushright}

\vspace{1cm}
\begin{center}
\begin{Large}
{\bf The chiral structure of the Lamb shift and 
the definition of the proton radius}\\[2cm] 
\end{Large} 
{\large Antonio Pineda}\footnote{pineda@ecm.ub.es}\\
{\it Dept. d'Estructura i Constituents de la Mat\`eria and IFAE,
U. Barcelona \\ Diagonal 647, E-08028 Barcelona, Catalonia, Spain}
\end{center}

\vspace{1cm}

\begin{abstract}
The standard definition of the electromagnetic radius of 
a charged particle (in particular the proton) is ambiguous once 
electromagnetic corrections are considered. We argue that a natural 
definition can be given within an effective field theory framework 
in terms of a matching coefficient. The definition of the neutron radius 
is also discussed. We elaborate on the effective field 
theory relevant for the hydrogen and muonic hydrogen, specially for the 
latter. We compute the hadronic corrections to the lamb shift 
(for the polarizability effects only with logarithmic accuracy) within 
heavy baryon effective theory. We find that they diverge in the inverse 
of the pion mass in the chiral limit.
\vspace{5mm} \\
\end{abstract}

\end{titlepage}
\vfill
\setcounter{footnote}{0} 
\vspace{1cm}

\section{Introduction}

The possibility
to measure the muonic hydrogen ($\mu p$) Lamb shift at the 30 ppm level at the PSI 
\cite{Taqqu} and,
consequently, to obtain the proton radius with an improved accuracy of, at
least, one order of magnitude better than current determinations, has triggered
 a lot of theoretical effort
in order the theoretical errors to compete with the estimated experimental
ones. We refer to \cite{pachucki1,pachucki2,pachucki4} and references therein for a
detailed explanation of the actual status of the art. On the other 
hand, it is also planed to study the muonic hydrogen properties at a 
future Neutrino Factory Complex \cite{CERN}. 

Measuring the proton radius with such accuracy poses some 
theoretical questions on the definition of the proton radius. Actually, 
these problems already appear with the present claimed precision of around 
1.5 \% (see for instance \cite{Karshenboim:1997zu}). 
The problem has to do with the fact that the proton radius is only 
defined at leading order in the electromagnetic coupling $\al$. Beyond 
this order, the usual definition through the use of form factors runs 
into problems because it becomes infrared divergent. This fact has 
already been discussed by Pachucki \cite{Pachucki3}. Here we address 
the problem using potential NRQED (pNRQED) \cite{Mont,lamb,pos}, 
an effective field theory suited for the study of non-relativistic systems. 
We will closely follow Ref. \cite{HF}, where the hyperfine splitting of the muonic 
hydrogen  
and the hydrogen ($ep$) was studied (we refer to this reference for further details, 
definitions and notation). Within 
this framework, and beyond leading order in $\al$, we will see that 
what one is really measuring is a matching coefficient of the effective theory,
which is dependent on the scale (scheme) at which the matching 
has been performed.  

On the other hand we will see that there is a series of hadronic 
contributions 
that can be computed using heavy baryon effective theory (HBET) \cite{HBET}. 
In particular,
the non-analytic (and model-independent) dependence on the light quark masses 
of the Lamb shift can be obtained. We will do so here for the Zemach and 
vacuum polarization 
piece of the hadronic correction. The polarizability corrections 
will only be considered with logarithmic accuracy. The full computation 
will be presented elsewhere.

\section{Effective Field Theories}

We will obtain pNRQED by
passing through different intermediate effective field theories after
integrating out different degrees of freedom. The path that we will
take is the following:
$$
{\rm HBET} \rightarrow ({\rm QED}) \rightarrow {\rm NRQED} \rightarrow
{\rm pNRQED},
$$
much in the same way as we did in Ref. \cite{HF}. The fact that we are 
interested in the Lamb shift will force us to consider other operators 
which were not considered in that paper. 

\subsection{HBET}
\label{secHBET}

Our starting point is the SU(2) version of HBET coupled to leptons,
where the delta is kept as an explicit degree of freedom.  The degrees
of freedom of this theory are the proton, neutron and delta, for which
the non-relativistic approximation can be taken, and pions, leptons
(muons and electrons) and photons, which will be taken relativistic.
This theory has a cut-off $\mu << m_p$,
$\Lambda_{\chi}$ and much larger than any other scale in the problem.
The Lagrangian can be split in several sectors. Most of them have
already been extensively studied in the literature but some will be
new.  Moreover, the fact that some particles will only enter through
loops, since only some specific final states are wanted, will simplify
the problem. The Lagrangian can be written as an expansion in $e$ and $1/m_p$.
and can be structured as follows
\be
\label{LHBET}
{\cal L}_{HBET}=
{\cal L}_{\gamma}
+
{\cal L}_{l}
+
{\cal L}_{\pi}
+
{\cal L}_{l\pi}
+
{\cal L}_{(N,\Delta)}
+
{\cal L}_{(N,\Delta)l}
+
{\cal L}_{(N,\Delta)\pi}
+
{\cal L}_{(N,\Delta)l\pi},
\ee
representing the different sectors of the theory. In particular, the
$\Delta$ stands for the spin 3/2 baryon multiplet (we also use
$\Delta=m_{\Delta}-m_p$, the specific meaning in each case should be
clear from the context).

Our
aim is to obtain the Lamb shift with
$O(m_{l_i}^3\al^5/m_p^2F(m_{l_i}/m_\pi,m_\pi/\Delta)$
accuracy. In particular, we want to obtain the non-analytic behavior in 
the masses of the light quarks and in the $\Delta$. 
Actually, the full dependence on the ratio 
$m_\pi/\Delta$ can be obtained using chiral Lagrangians. 
In this paper we will mainly focus on a partial contribution 
to it: the so-called Zemach correction. 
 
To obtain this accuracy, we need,
in principle, Eq. (\ref{LHBET}) with $O(1/m_p^2)$ accuracy. The different 
pieces of the Lagrangian can be found in Ref. \cite{HF}. There are a few 
differences due to the fact that we are now concerned with the Lamb 
shift correction. This implies that some terms in the nucleon Lagrangian, 
which were not considered in Ref. \cite{HF} have to be considered here. 
In particular, the term 
\be
\delta {\cal L}_{\gamma}=\left({d_{2,R} \over m_p^2}+{d_{2}^{(\tau)} \over m_{\tau}^2} 
\right) F_{\mu \nu} D^2 F^{\mu \nu}
\label{phm2}
\,,
\ee
where $d_{2,R}$ stands for the hadronic contribution. 
This term can be eliminated by field redefinitions (see for instance 
\cite{Pineda:2000sz}) but we will 
keep it explicit since this contribution can be singled out by considering the 
lamb shift of a purely leptonic bound state. 

For the nucleon there is a new term that has to be considered (actually in this 
one is where the proton radius is encoded):
\be
\label{LNdelta}
\delta {\cal L}_{N}= N^\dagger_{p} \Biggl\{
 -e{c_D^{(p)} \over m_p^2}
   \left[{\bf \bfnabla \cdot E }\right] 
\Biggr\} N_{p}
\,.
\ee

Specially relevant in our case is $c_{3,R}^{pl_i}$ (see Eq. (14) in \cite{HF}). 
Nevertheless, the leading contribution in $\al$ (of $O(\al^2)$) is suppressed by 
an extra factor $m_{l_i}/m_p$ (see Eq. (\ref{c3})), i.e. $c_{3,R}^{pl_i} 
\sim \al^2 m_{l_i}/m_p$, which 
goes beyond the aimed accuracy of our calculation and we can neglect it.

\subsection{NRQED}

Since we want to obtain corrections to the lamb shift, besides the 
operators considered in Ref \cite{HF} new operators have to be 
considered with $O(1/(m_p^2m_\pi))$ precision. This means that some operators 
of up to dimension 7 have to be included in the Lagrangian. They read

\be
\label{LNdeltaNR}
\delta {\cal L}_{N}= N^\dagger_{p} \Biggl\{
  c^{(p)}_{A_1}\, {e^2}\, {{\bf B}^2-{\bf E}^2 \over 8 m_p^3}
- c^{(p)}_{A_2}\, {e^2}\, {{\bf E}^2 \over 16 m_p^3} 
\Biggr\} N_{p}
\,.
\ee
$c_{A_1}^{(p)}$, $c_{A_2}^{(p)}$
correspond to the proton polarizabilites. The relations
are the following:
\be
c_{A_1}^{(p)}=4m_p^3{\beta_M^{(p)} \over \alpha} \,,
\qquad
c_{A_2}^{(p)}=-{8m_p^3\over  \alpha}(\alpha_E^{(p)}+\beta_M^{(p)}).
\ee

We also note that the operators considered in Eqs. (\ref{phm2}) and 
(\ref{LNdelta}) also appear here. In the first case $d_{2,R} \rightarrow 
d_{2,NR}$. $d_{2,NR}$ is scale independent 
but pion effects should be included now. We will discuss in further 
detail these corrections in sec. \ref{sec:matching}. 
$c_D^{(p)}(\nu)$ is a matching-scale dependent quantity and 
has already been considered in HBET. The difference is the 
scale $\nu$ considered in each case, whereas in HBET 
$m_{\pi} \ll \nu \ll m_p$, we now have $m_{l_i}\al \ll \nu \ll m_{\pi}$. 
This means that in HBET, effects due to the pion scale were not 
included in $c_D^{(p)}(\nu)$ but now they are. A similar discussion 
applies to $c_{3,NR}^{pl_i}(\nu)$ (see Eq. (17) in Ref. \cite{HF}), 
which is also scale dependent.

\subsection{pNRQED}

After integrating out scales of $O(m_{l_i}\al)$ one ends up in a
Schr\"odinger-like formulation of the bound-state problem. We refer to
\cite{Mont,lamb,pos} for details.  The pNRQED Lagrangian for the $ep$ (the
non-equal mass case) can be found in Appendix B of \cite{pos} up to $O(m\al^5)$. The
pNRQED Lagrangian for the $\mu p$ is similar except for the fact that
light fermion (electron) effects have to be taken into account. The explicit 
form of the Lagrangian has not been presented so far. We do so here. The
explicit Lagrangian reads (up to $O(m\al^5)$)
\bea
&&L_{pNRQED} =
\int d^3{\bf x} d^3{\bf X} S^{\dagger}({\bf x}, {\bf X}, t)
                \Biggl\{
i\partial_0 - { {\bf p}^2 \over \mu_{12}} + { {\bf p}^4 \over 8m_{\mu}^3}+ 
{ {\bf p}^4 \over 8m_p^3} - { {\bf P}^2 \over 2M}
\\
&&
\nonumber
- V ({\bf x}, {\bf p}, {\bfsigma}_1,{\bfsigma}_2) + e {\bf x} \cdot {\bf E} ({\bf X},t)
\Biggr\}
S ({\bf x}, {\bf X}, t)- \int d^3{\bf X} {1\over 4} F_{\mu \nu} F^{\mu \nu}
\,,
\eea
where $M=m_{\mu}+m_p$, $\mu_{12}= {m_{\mu}m_p \over m_{\mu}+m_p}$, ${\bf x}$ and 
${\bf X}$, and
${\bf p}$ and ${\bf P}$ are the relative and center of mass coordinate and momentum
respectively.

$V$ can be written as an expansion in $1/m_{\mu}$, $1/m_p$, $\al$, ... 
We will assume $1/r \sim m_e$ (which is realistic for the case at hand) 
and concentrate in the $1/m_{\mu}$ expansion. 
We assume that $m_{\mu} << m_p$.
\be
V ({\bf x}, {\bf p}, {\bfsigma}_1,{\bfsigma}_2)
=
V^{(0)}(r)+{V^{(1)}(r) \over m_{\mu}}+{V^{(2)}(r) \over m_{\mu}^2}+\,.\,.\,.
\ee
In order to reach the desired NNNLO accuracy in $\als$, $V^{(0)}$ 
has to be computed up to $O(\al^4)$, $V^{(1)}$ up to $O(\al^3)$, 
$V^{(2)}$ up to $O(\al^2)$ and $V^{(3)}$ up to $O(\al)$. Nevertheless, 
there is not contribution to $V^{(3)}$ at tree level.
We first write the potential up to order $1/m_{\mu}^2$ in momentum space.\\
{\bf Order $1/m^0$}. We define ${\tilde V}^{(0)}$ 
(the Fourier transform of $V^{(0)}$):
\begin{equation}
{\tilde V}^{(0)} \equiv  - 4\pi Z_{\mu}Z_p\alpha_{V}(k){1 \over {\bf k}^2},  
\label{defpot0}
\end{equation}
which in fact defines $\alpha_{V}$, which is gauge invariant. 
There is another very popular definition that enjoys all the nice properties
we would like $\al$ to have (gauge invariance,scheme/scale independence...):
\be
\alpha_{eff}(k)=\al{1 \over 1+\Pi(-{\bf k}^2)}
\,,
\ee
where 
$$
\Pi(k^2)=\al\Pi^{(1)}(k^2)+\al^2\Pi^{(2)}(k^2)+\al^3\Pi^{(3)}(k^2)+...
$$ 
is the electron vacuum polarization and $\Pi(0)=0$. 
$\alpha_{eff}$ corresponds to Dyson summation. If we express
$\alpha_{V}(k)$ in terms of $\alpha_{eff}(k)$, we have 
\be 
\alpha_{V}(k)=
\alpha_{eff}(k)+\sum_{n,m=0 \atop n+m=even>0} Z_{\mu}^nZ_p^m\al_{eff}^{(n,m)}(k) 
=\alpha_{eff}(k)+\delta \al(k)\,, \quad \quad \delta \al(k)=O(\al^4)
.
\ee
The sum is constrained to fulfill $n+m=even$ because of the Furry theorem.
Each $\al_{eff}^{(n,m)}(k)$ is also gauge invariant. 
In order to achieve $O(m\al^5)$ accuracy we need to know
$\Pi^{(1)}$, $\Pi^{(2)}$, $\Pi^{(3)}$ (see \cite{Kinoshita:1979dy}) 
and the leading, non-vanishing,
contributions to $\al_{eff}^{(2,0)}(k)$, $\al_{eff}^{(0,2)}$ and
$\al_{eff}^{(1,1)}$. A discussion on the latter (which still remain to be 
computed) can be found in Ref. \cite{Borie:1982ax}.

As we have already mentioned, the matching is performed in a expansion in
$1/m$, $\al$ and the energy. The leading-non-zero contribution to the
potential from the expansion in the energy appears at
$O(1/m^0)$ and $O(\al^2)$ and it will give rise to a contribution of
$O(m\al^5)$. It follows from the electron vacuum polarization correction 
to the potential shown in Fig. 1. In momentum space it reads 
\be
\delta {\tilde V}_{VP} = Z_{\mu}Z_pe^2{1 \over {\bf k}^2} \pi(k^2)
\,,
\ee
where $\pi(k^2)=\al\Pi^{(1)}(k^2)$. 
We note that this correction depends on $k_0$. Since $k_0
\sim m\al^2$ and ${\bf k} \sim m\al$ we can expand in ${k^{2}_0 \over {\bf
    k}^2}$. The leading order is $k_0$ independent and produces the standard
vacuum polarization correction that we have already taken into account in 
$\al_{eff}$:
\be
Z_{\mu}Z_pe^2{1 \over {\bf k}^2} \pi(-{\bf k}^2)
\,.
\ee
The next correction reads
\be
\delta {\tilde V}_E = Z_{\mu}Z_pe^2{ k^{2}_0 \over {\bf k}^2} \pi^{\prime}(-{\bf k}^2)
\label{potener}
\,,
\ee
and is new.

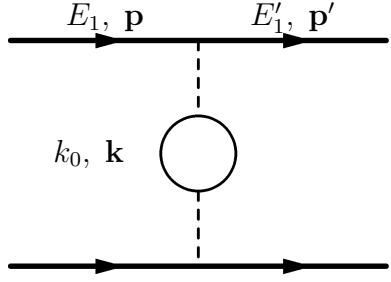
\begin{figure}[!htb]
\begin{center} 
  \begin{fmfgraph*}(50,30)
       \fmfstraight
       \fmftop{i1,o1}
       \fmfbottom{i2,o2}
       \fmf{fermion,width=thick,label=$E_1,,\;{\bf p}$,label.side=left}{i1,v1}
       \fmf{fermion,width=thick,label=$E_1^{\prime},,\;{\bf
       p}^{\prime}$, label.side=left}{v1,o1}
       \fmf{fermion,width=thick}{i2,v2,o2}
       \fmffreeze
       \fmf{dashes,width=thin}{v1,v3,v2}
       \fmfv{l=$k_0,, \; {\bf k}$,l.a=180,label.dist=.2w,decoration.shape=circle,d.f=empty,d.si=.2w}{v3}
    \end{fmfgraph*}
\end{center}

\vspace{0.3cm}
\caption{\it Leading correction to the Coulomb potential due to the electron vacuum
  polarization in the Coulomb gauge. ${\bf k}={\bf p}-{\bf p}^{\prime}$ and 
  $k_0=E_1-E_1^{\prime}$.}
\label{figvacuum}
\end{figure}

The explicit expression of $\pi$ reads
\be
\pi (k^2) = {\al \over \pi} k^2 \int_4^{\infty}d(q^2){1 \over
  q^2(m_e^2q^2-k^2)}u(q^2)
\,,
\ee
where
$$
u(q^2)={1 \over 3}\sqrt{1-{4 \over q^2}}\left(1+{2 \over q^2}\right)
\,.
$$
The $k_0$ dependence in the potential translates into terms with time
derivatives in the pNRQED Lagrangian (or in other words an energy dependent
potential). At the order we are working, one can get rid of these time derivatives by using the equations
of motion. Note that we refer to the full equations of motion (including the
potential), and not just the equations of motion of free particles, reflecting
the fact that the matching has to be done off-shell. The
procedure has been illustrated in the Appendix A of  
\cite{pos}. Effectively, this
transforms $\delta {\tilde V}_E$ into the following expression 
\be
\delta {\tilde V}_E = -{Z_{\mu}Z_pe^2 \over 4 m_{\mu}m_p}
{({\bf p}^2-{\bf p}^{\prime\,2})^2 \over {\bf k}^2} {\al \over \pi} 
m_e^2 \int_4^{\infty}d(q^2){1 \over
  (m_e^2q^2+{\bf k}^2)^2}u(q^2)
\,.
\label{dtv}
\ee

We see that it becomes a $O(1/m^2)$ potential. Field redefinitions should 
produce an equivalent result. We note that if we want to go to higher
orders than $O(m\al^5)$, we should work with the former expression
Eq. (\ref{potener}), or work in a more general way by using field redefinitions.

{\bf Order $1/m$}. They will give rise to $O(m\al^6)$ corrections so they will
be neglected in the following.

{\bf Order $1/m^2$}. At the order we are working they can be written as
follows\footnote{We have slightly changed the definitions with respect 
\cite{HF}. We now set $Z_{l_i}=Z_p=1$ in the terms proportional to non-trivial 
matching coefficients in the Lagrangian (ie. $c_F$, $c_S$, $\cdots$). This allows 
to deduce the neutron case by setting $Z_p \rightarrow 0$ afterwards.}: 
\bea
&&
\label{VbcD}
{\tilde V}^{(b)} = { \pi\al_{eff}(k) \over 2} \left[
  Z_p{c_D^{(\mu)} \over m_{\mu}^2}+Z_{\mu}{c_D^{(p)} \over m_p^2}
\right]
\,,
\\
&&
{\tilde V}^{(c)} = -  i 2 \pi\al_{eff}(k)  { ({\bf p} \times
  {\bf k})  \over {\bf k}^2}\cdot \left\{ Z_p{c_S^{(\mu)} {\bf s}_1 \over m_{\mu}^2} + 
Z_{\mu}{c_S^{(p)} {\bf s}_2 \over m_p^2}
\right\}
\,,
\\
&&
{\tilde V}^{(d)} =- Z_{\mu}Z_p 16\pi\al\left({ d_2^{(\mu)} \over m_{\mu}^2}
+{ d_2^{(\tau)} \over m_{\tau}^2}+{ d_{2,NR} \over m_p^2}
\right)
\,,
\\
&&
{\tilde V}^{(e)} = - Z_{\mu}Z_p{ 4 \pi\al_{eff}(k) \over m_{\mu}m_p} \left( { {\bf p}^2
  \over {\bf k}^2} - {({\bf p} \cdot {\bf k})^2 \over {\bf k}^4}
\right)
\,,
\\
&&
{\tilde V}^{(f)} = - { i 4 \pi\al_{eff}(k) \over m_{\mu}m_p} { ({\bf p} \times
  {\bf k})  \over {\bf k}^2}\cdot(Z_pc_F^{(\mu)} {\bf s}_1+Z_{\mu}c_F^{(p)} {\bf s}_2)
\,,
\\
&&
{\tilde V}^{(g)} = { 4 \pi\al_{eff}(k) c_F^{(1)}c_F^{(2)} \over m_{\mu}m_p} 
\left( {\bf s}_1  \cdot {\bf s}_2 - { {\bf s}_1 \cdot {\bf k} {\bf
      s}_2 \cdot {\bf k} \over {\bf k}^2} 
\right)
\,,
\\
&&
{\tilde V}^{(h)} = -{1 \over m_{p}^2}
\left\{(c_{3,NR}^{pl_i}+3c_{4,NR}^{pl_i}) 
-2c_{4,NR}^{pl_i} {\bf S}^2 \right\}
\,.
\eea
The classification of the diagrams follows the one of Fig. 1 
of Ref. \cite{pos} generalized to the non-equal mass case and with the replacement  
$\al \rightarrow \al_{eff}$. 
This ensures that the expressions for the potentials are correct at one-loop (the 
one-loop vacuum polarization is included). Written in this way subleading terms in 
$\al$ are also 
included. It should be keep in mind that at some point the replacement of 
$\al$ by $\al_{eff}$ is not enough and new corrections have to be included in the 
potentials.
\be
d_{2,NR}=d_{2,R}+{m_p^2 \over4}\Pi^\prime_{h,\pi}(0)={m_p^2 \over4}\Pi^\prime_{h}(0)
\,,
\ee
where $\Pi^\prime_{h}(0)$ is the derivative of the hadronic vacuum polarization
(we have defined $\Pi_{h}(-{\bf k}^2)=-{\bf k}^2\Pi^\prime_{h}(0)+\,.\,.\,.$) 
and $\Pi^\prime_{h,\pi}(0)$ is the contribution due only to pions. 

The one loop diagrams of fig. 2 of Ref. \cite{pos} are equal in our case up 
to a trivial mass rescaling:
\be
{\tilde V}^{(a)}_{1loop} = { Z_{\mu}^2Z_p^2 \al^2 \over m_{\mu}m_p} \left(\log {{\bf k}^2
    \over \mu^2} - {8 \over 3}\log2 + {5 \over 3} \right)
\,,
\ee
\be
{\tilde V}^{(b,c)}_{1loop} = { 4 Z_{\mu}^2Z_p^2 \al^2 \over 3m_{\mu}m_p} \left(\log {{\bf k}^2
    \over \mu^2} + 2 \log2 - 1 \right)
\,.
\ee

The procedure we have followed to obtain the potentials is different from the one 
used in Ref. \cite{pachucki1}. For instance, in this last reference no mention to 
a correction to the type of Eq. (\ref{dtv}) is made. To facilitate the 
comparison one should work in position space (it is also eventually useful to write 
the potential in a basis where the angular momentum structure of the potential is 
more visible). The non-trivial comparison can be reduced to the contribution 
due to $\delta {\tilde V}_E$ and the one-loop correction to ${\tilde V}^{(e)}$, 
which in our case reads
\bea
&&
\left. 
\delta V_E +V^{(e)}\right|_{1 \rm loop} = -{Z_{\mu}Z_p\al^2 \over \pi}{m_e^2 \over
  m_{\mu}m_p}\int_4^{\infty}d(q^2){u(q^2) \over (m_eq)^2}
\nn
\left\{
{1 \over 2}\left\{{\bf p}^2,{e^{-m_eqr} \over r}
\left(1+{m_eqr \over 2}\right)\right] \right\}
\nn
\\
&&
-{e^{-m_eqr} \over 2r^3}\left(1+m_eqr \right){\bf L}^2
\left.
+
{(m_eq)^2 \over 4r}e^{-m_eqr}\left(1+{m_eqr \over 2}\right)-2\pi\delta(\bf r)
\right\}
\,,
\label{potential}
\eea
where ${\bf L}$ is the angular momentum. We have checked that this expression 
agrees with the corresponding expression in Ref. \cite{pachucki1}.

\medskip

In this paper, our main focus are the hadronic 
corrections, in particular those proportional to the $\delta(\bf r)$ 
potential\footnote{There are other hadronic corrections. In particular, 
one also has to 
consider the one-loop vacuum-polarization correction in Eq. (\ref{VbcD}), 
proportional to $c_D^{(p)}$. Nevertheless, this term has a different functionality 
on $r$, which makes it possible (at least as a gedanken experiment) to single 
it out by considering different observables (the result will depend on 
$m_{\mu}\al/(nm_e)$ for the spectrum or on $|{\bf p}|/m_e$ above threshold). 
Therefore, we will not consider it in this paper although it should be 
considered for the Lamb shift of low excitations of muonic hydrogen 
with $O(m\al^5)$ accuracy (see \cite{pachucki1} for an explicit computation).}. 
They can be encoded in the matching coefficient of 
 the spin-independent delta potential:
\be
\delta V = { D_{d}^{\rm had} \over m_p^2}\delta^{(3)}({\bf r})
\,.
\ee
where
\be
D_{d}^{\rm had} \equiv -c_{3,NR}^{pl_i}-16\pi\al Z_{l_i}Z_pd_{2,NR}+
{\pi\al \over 2}Z_{l_i}c_D^{(p)}
\,.
\ee
The spin-dependent term does not contribute to the average energy  
over polarizations. Therefore the hadronic contribution to the lamb shift reads
\be
\delta E=\langle E(s)-E(p) \rangle ={ D_{d}^{\rm had} \over m_p^2}\delta_{l0}{1 \over \pi}
\left(
{\mu_{l_ip}\al \over n}
\right)^3
\,.
\ee

\section{Definition of the proton radius}

The usual definition of the electromagnetic radius of the proton is the following
\be
\label{rpGpE}
r^2_p=6 \left.{d G_{p,E}(q^2) \over d\,q^2}\right|_{q^2=0}
\,.
\ee
This definition runs into problem once electromagnetic corrections are included in 
the computation of $G_{p,E}$
because it becomes infrared divergent (this is so for a charged particle, 
for the neutron this definition is infrared safe). 
This problem has already been pointed out in Ref. \cite{pachucki1}. Nevertheless, 
this is not a problem from the effective theory point of view, where the parameters 
of the theory (the matching coefficients) do depend in general 
on the scheme and scale on which the computation has been performed. 
Therefore, our definition of the proton radius is the following
\be
\label{cDrp}
c_D^{(p)}-1 \equiv {4\over 3}r_p^2 m_p^2
\,,
\ee
which naturally relates to the one in Eq. (\ref{rpGpE}), since 
\be
c_D=1+2F_2+8F_1^{\prime}=1+8m_p^2\left.{d G_{p,E}(q^2) \over d\,q^2}\right|_{q^2=0}
\,,
\ee
where we have already used that $F_1=1$ due to gauge invariance. 
Eq. (\ref{cDrp}) is the natural definition from the point of view of effective field 
theory (a matching coefficient). Another issue is the separation of  
"hadronic" from electromagnetic corrections. 
We stress that the electromagnetic effects are included in the form factors above.
We note that $r_p$ is now an scale dependent quantity and fulfils an expansion 
in $\al$:
\be
r_p(\nu)=r_p^{(0)}+\al r_p^{(1)}(\nu)+\cdots
\,.
\ee
We will define the pure hadronic definition of the radius 
("hadronic proton electromagnetic radius") 
to be the first term in this expansion, $r_p^{(0)}$. Note that this 
quantity is not an observable from the effective field theory point of view, since 
it will always appear (in the same combination) with another hadronic corrections 
in the observable. This definition for the "hadronic" piece  
is different in general\footnote{To get 
a closer connection between both definitions one should take $\mu_1< m_e$ in 
Ref. \cite{Gasser:2003hk}, where $\mu_1$ is the matching scale at which the parameters 
of the theory are make equal between the hadronic and hadronic+electromagnetic case. 
In other words, one natural place to make the connection between 
hadronic and hadronic+electromagnetic quantities is at very low energies
($\mu_1 \rightarrow 0$), where QED has an infrared fixed point and consequently the 
fundamental parameters of QCD do not run.} 
of the definition in \cite{Gasser:2003hk}.

We can also know what the running of each term of $r_p$ is. As we have already 
mentioned, the first term is scale independent but not the second (even if this 
running can be understood in terms of point-like particles).
\be
\nu{ d \over d \nu} r_p^{(0)}=0\,, \quad 
r_p^{(0)}\nu{d \over d \nu} r_p^{(1)}=-{1 \over \pi}{1 \over m_p^2}\,,\quad \cdots
\,.
\ee

In summary, if we give a number for the proton radius we have to say in which 
scheme (for instance $\MS$) and scale it has been obtained if a precision of 
around 1 \% is claimed. In Ref. \cite{Pachucki3} a prescription has been taken to 
fix the ambiguity with $O(\al)$ precision (how to generalize this prescription to 
higher orders in $\al$ is left unsettled). There, the scale dependence
at $O(\al)$ is regulated by the proton mass and the finite pieces fixed 
by performing the computation assuming the proton to be point-like at scales 
of order $m_p$. Nevertheless, this assumption is not justified  
since at these scales several resonances appear. 
We would also like to stress that our definition of the proton radius through 
a matching coefficient, $c_D^{(p)}$, makes easier to obtain the dependence of 
any observable in this quantity, in particular for the cross section 
$\sigma(e^- p \rightarrow e^- p)$. 
  
\medskip
 
What we have said is quite general for any nucleus. Indeed, 
the object which is really measured at very low 
momentum transfer (above or below threshold) 
between the nucleus and lepton is $D_d(\nu)=D_{d}^{\rm lep}+D_{d}^{\rm had}$. 
Therefore, one could think that it is more proper to give numbers for this object 
(for a given scale and scheme).
Nevertheless, if we want to get some 
understanding on the structure of the nucleus it maybe convenient to 
distinguish the different hadronic and leptonic contributions of $D_d(\nu)$ 
as some of them 
can be singled out from different measurements. For instance the contribution due to 
$d_{2,NR}$, as well as 
the  leptonic contributions, can be singled out since they can be computed explicitly 
and measured changing the 
nucleus by another lepton. Moreover, it also makes some sense to distinguish between 
$c_{3,NR}^{pl_i}$ and $c_D^{(p)}$ (at least at $O(\al^2m_{l_i})$), since 
the leading dependence on $c_{3,NR}^{pl_i}$ is proportional to the lepton 
mass and can be distinguished by changing the lepton. Beyond $\al^2$ or beyond 
$m_{l_i}$, this 
distinction is not that clear and can not be achieved in general. This has to 
do with the fact that we are not working in a minimal basis of operators in the 
effective Lagrangian. In other words, one can always perform a field redefinition 
to eliminate $c_D^{(p)}$, which gets absorbed in $c_{3,NR}$. Therefore, in the long 
term one should work with $D_d^{\rm had}$, 
which can be related with observable quantities (for the neutron case it can easily 
be related with the scattering lengths\footnote{The definition of the neutron radius 
could also be worked out on similar terms. 
One could write the effective Lagrangian, which would be similar to the proton one 
but for a particle with zero charge, i.e. with the replacement of covariant by 
normal derivatives. The definition of the neutron radius is also 
\be
r^2_n=6 \left.{d G_{n,E}(q^2) \over d\,q^2}\right|_{q^2=0}
\,,
\ee
and from the effective field theory point of view
\be
\label{cDrn}
c_D^{(n)} \equiv {4\over 3}r_n^2 m_n^2
=2F_2^{(n)}+8F_1^{(n)\prime}
\,,
\ee
where we have already used that $F_1^{(n)}=0$ due to gauge invariance. 
The point that we would like to remark here is that $r_n^2$ does not correspond 
to the neutron scattering length once $O(\al)$ corrections are included. Rather the 
neutron scattering length $b_{n l_i}$ (where $l_i$ represents the specific lepton) 
reads (note that $d_{2,NR}$ does not contribute in this case)
\be
b_{nl_i}={1 \over 4m_n}\left( \al Z_{l_i}c_D^{(n)} -{2 \over \pi}c_{3,NR}^{nl_i}\right)
\,,
\ee
This is a low-energy constant (formally we can take the limit of transfer momentum 
going to zero) and measurable quantity which can (in principle) 
be obtained with arbitrary accuracy from experiment. Note also that if 
one works in a minimal basis of operators in the effective theory (by absorbing  
$c_D^{(n)}$ in $c_{3,NR}^{nl_i}$ through field redefinitions) there is a one-to-one 
correspondence of the scattering length with a matching coefficient of the 
effective theory: $c_{3,NR}^{nl_i}$.}).

\section{Matching HBET to NRQED}
\label{sec:matching}

The hadronic corrections to the spin-independent delta potential are encoded in the 
corrections to $d_{2,NR}$, $c_D^{(p)}$ and $c_{3,NR}$. 

We will only compute the hadronic corrections to $d_{2,NR}$ and $c_{3,NR}$ (due to
 energies 
of $O(m_{\pi},\Delta)$), since our aim is to obtain $c_D^{(p)}$ from 
experiment (for instance, from the muonic hydrogen lamb shift). 
If we were interested in $r_p^{(0)}$ one should also compute the hadronic 
$O(\al)$ corrections to $c_D^{(p)}$.

The leading correction to $d_{2,NR}$ is due to the one-loop pion correction to 
the vacuum polarization. The diagram that contributes at leading order 
is the same than in Fig.  \ref{figvacuum} 
changing the electron by a pion in the limit of transfer momentum going to zero. 
It reads
\be
\Pi^\prime_{h,\pi}(0)={\al \over 30\pi}{1 \over 4m_{\pi}^2 }
\,.
\ee

\begin{figure}[h]
\makebox[2.0cm]{\phantom b}
\epsfxsize=9truecm \epsfbox{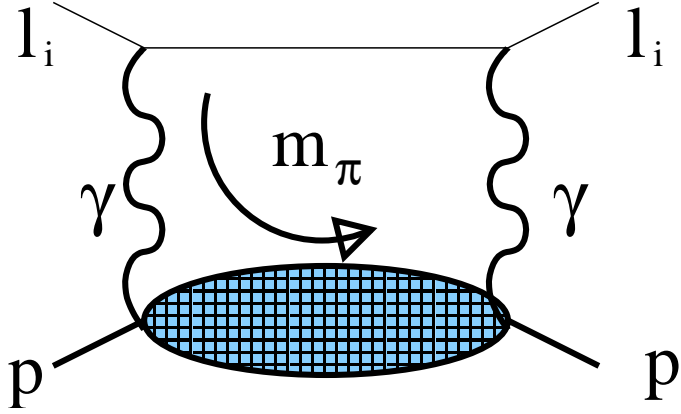}
\caption {{\it Symbolic representation of Eq. (\ref{c3}).}}
\label{figc3}
\end{figure}

The contribution to $c_{3,NR}$ (see Fig. \ref{figc3}) 
can be written in a compact way in 
terms of the form factors of the forward virtual-photon Compton 
tensor (see \cite{HF} for details). It reads (this expression is 
only valid in 4 dimensions and in $D$-dimensions has to be properly generalized)
\be
\label{c3}
c_{3,NR}^{pl_i}=i g^4 m_pm_{l_i}\int {d^4k \over (2\pi)^4}{1 \over k^4}{1 \over
k^4-4m_{l_i}^2k_0^2 }
\left\{
S_1(k_0,k^2)(-3k_0^2+{\bf k}^2)-{\bf k}^2S_2(k_0,k^2)
\right\}
\,.
\ee
This result keeps the complete dependence on $m_{l_i}$. This contribution 
is usually organized in the following way
\be
\label{c3split}
c_{3,NR}^{pl_i}=
c_{3,R}^{p}
+\delta c_{3,point-like}^{pl_i}
+\delta c_{3,Zemach}^{pl_i}+\delta c_{3,pol.}^{pl_i}
\,.
\ee
We remind that $c_{3,R}^{p}$ is $O(\al^2 m_{l_i}/m_p)$ and can be neglected. 
The second term corresponds to the diagram assuming the proton non-relativistic 
and point-like. With the precision needed it reads (in the $\MS$ scheme)
\be
\label{c3pointlike}
\delta c_{3,point-like}^{pl_i}
={m_p \over m_{l_i}}\left(\ln{m_{l_i}^2 \over \nu^2}+{1 \over 3}\right)
\,.
\ee


\begin{figure}[h]
\makebox[2.0cm]{\phantom b}
\epsfxsize=9truecm \epsfbox{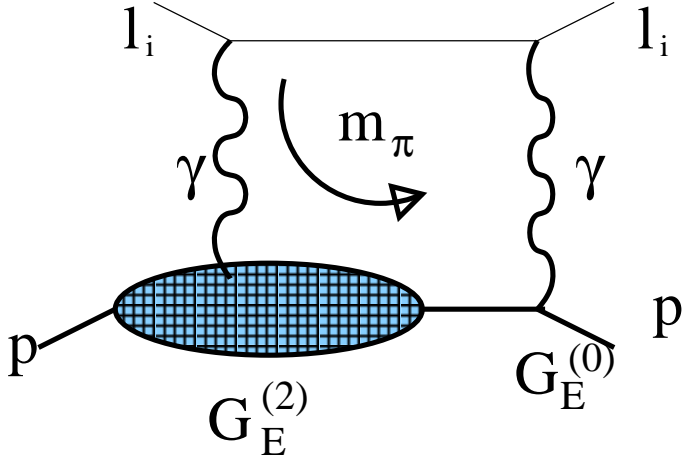}
\caption {{\it Symbolic representation (plus permutations) 
of the Zemach correction in Eq. (\ref{c3Zemach}).}}
\label{figzemach}
\end{figure}

The second term in Eq. (\ref{c3split}) is the so-called Zemach correction, which is 
one of the main concerns of this paper. It is symbolically pictured in 
Fig. \ref{figzemach} and reads
\be
\label{c3Zemach}
\delta c_{3,Zemach}^{pl_i}= 
4(4\pi\alpha)^2m_p^2m_{l_i}\int {d^{D-1}k \over (2\pi )^{D-1}}
{1 \over {\bf k}^6}G_E^{(0)}G_E^{(2)}
\,.
\ee
This term appears to be finite to the order of interest (it produces
no logs) and it agrees with the result obtained by Pachucki \cite{pachucki1}
at leading order. Note, that this result holds for both ep and $\mu p$. In other 
words, the exact dependence on $m_{l_i}$ is kept. $G_E^{(0)}=1$. We take the 
expression for $G_E^{(2)}$ from \cite{BFHM}. 
The use of effective field theories and dimensional regularization 
is a strong simplification: we only need the non-analytic behavior of $G_E^{(2)}$ 
in $q^2$, the analytical behavior produces scaleless integrals, which are zero 
in dimensional regularization. This is a reflection of the factorization of the 
different scales. We do not need to introduce the point-like interactions to 
regulate the infrared divergences of the integrals at zero momentum. A similar 
thing happened \cite{HF} for the computation of the hyperfine splitting 
of the hydrogen, where, again, there was not not need to introduce the 
point-like interactions to regulate the infrared divergences of the 
integrals at zero momentum. 

The result of the computation reads
\bea
\delta c_{3,Zemach}^{pl_i}&=& 
{2}(\pi\al)^2 
\left({m_p \over 4\pi F_0 }\right)^2
{m_{l_i} \over m_\pi}
\left\{
{3 \over 4}g_A^2+{1 \over 8}
\right.
\\
&&
\left.
+{2 \over \pi}g_{\pi N\Delta}^2{m_\pi \over \Delta}
\sum_{r=0}^{\infty}C_r\left({m_\pi \over \Delta}\right)^{2r}
+g_{\pi N\Delta}^2\sum_{r=1}^{\infty}
H_r\left({m_\pi \over \Delta}\right)^{2r}
\right\}
\,,
\nn
\eea
where
\be
C_r={(-1)^r\Gamma(-3/2) \over \Gamma(r+1)\Gamma(-3/2-r)}
\left\{
B_{6+2r}-{2(r+2) \over 3+2r}B_{4+2r} 
\right\}
\,,
\qquad r \geq 0
\,,
\ee
and the definition of $B_n$ and $H_n$ can be found in \ref{app:formulas}. 
The term proportional to $B_n$ are due to scales of
$O(\Delta)$, the contributions proportional to $H_n$ are due to scales of
$O(m_\pi)$. We see that the latter give contributions to the Lamb
shift which are non-analytical in the quark
mass. 

We have then obtained the leading correction to the Zemach term in 
the chiral counting (supplemented with a large $N_c$ counting). This 
is a model independent result.  Other contributions to the Zemach term 
are suppressed in the 
chiral counting.

\medskip

\begin{figure}[h]
\makebox[2.0cm]{\phantom b}
\epsfxsize=9truecm \epsfbox{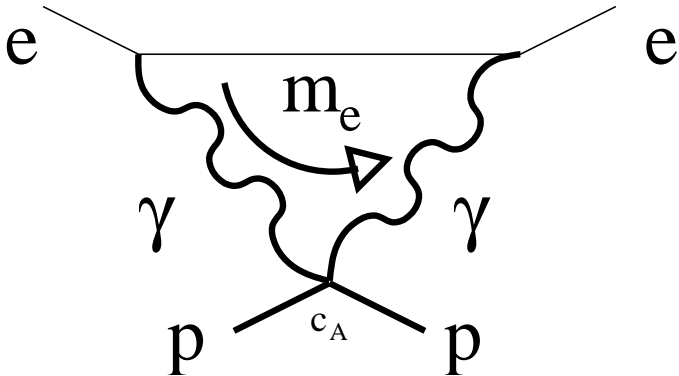}
\caption {{\it Diagram contributing to the
polarizability correction with $\ln m_e$ accuracy. The matching coefficients 
of the proton can be $c_{A_1}$ or $c_{A_2}$, or, in other words, the 
proton polarizabilities.}}
\label{figpolhyd}
\end{figure}

Finally, we consider the polarizability corrections. For those we would need 
the full expression for the virtual-photon Compton tensor at one-loop 
in HBET, which is lacking at present. 
This computation is relegated to a future publication. We will 
only consider here the limit $m_{l_i} \ll m_{\pi}$, which is relevant for the 
hydrogen atom, and only consider the logarithm ($\ln m_e$) enhanced terms. 
These can be obtained by computing the ultraviolet behavior of the diagram 
in Fig. \ref{figpolhyd}. This contribution is proportional to $c_{A_1}$ and $c_{A_2}$. 
It can be written in 
term of the polarizabilites of the proton (see \cite{Friar:1997tr,KS})
\be
\delta c_{3,pol}^{pl_i}=-\al m_p^2m_{l_i}\left[5\alpha_{E}^{(p)}-\beta_{M}^{(p)}\right]
\ln({m_{l_i}})
\,.
\ee
For the pure pion cloud, the polarizabilities were computed in Ref. 
\cite{Bernard:1992qa}. 
The contribution due to the $\Delta$ can be found in Ref. \cite{Hemmert:1999pz}.
The scale in the logarithm is compensated by the next scale of the problem, 
which can be $m_{\pi}$ or $\Delta$. For contributions which are only due 
to the $\Delta$ or pions, the scale is unambigous. In the case where pions 
and $\Delta$ are both present in the loop we will choose the pion mass (the 
difference being beyond the logarithmic accuracy). The result reads
\bea
\label{c3log}
\delta c_{3,pol}^{pl_i}&=&-{2 \over 9}\al^2 
{m_{l_i} \over \Delta}
b_{1,F}^2\ln{\Delta \over m_{l_i}}+{49 \over 12}\pi\al^2 g_A^2{m_l\over m_{\pi}} {m_p^2 \over 
(4\pi F_0)^2}
\ln({m_{\pi} \over m_l})
\\
\nn
&&
+{8 \over 27}\al^2 g_{\pi N\Delta}^2{m_l\over \sqrt{\Delta^2-m_{\pi}^2}} {m_p^2 \over 
(4\pi F_0)^2}
\left({45 \Delta \over \sqrt{\Delta^2-m_{\pi}^2}}
+{4\Delta^2-49m_{\pi}^2 \over \Delta^2-m_{\pi}^2}\ln R\right)\ln({m_{\pi} \over m_l})
\,,
\eea
where 
\be
R={\Delta \over m_{\pi}}+\sqrt{{\Delta^2 \over m_{\pi}^2}-1}
\,.
\ee
We would like to stress here that some of the corrections considered in Eq. 
(\ref{c3log}) are 
missing in the dipole approximation \cite{BE}, which is sometimes used in the 
literature. For instance, the first term in Eq. (\ref{c3log}) 
(the pure Delta correction) 
would be missing in that approximation. 

\section{Phenomenological analysis}

Our aim here is to see how much of the hadronic correction can be understood 
in terms of the physics of the $m_{\pi}$ and/or $\Delta$ scales.

There have been some previous estimates \cite{pachucki1,pachucki2,FM} 
of the Zemach correction using phenomenological model parameterizations of the 
proton form factors. They obtain quite similar numbers. For 
instance, using the dipole parameterization of the proton form factor,
\be
G_E({\bf q}^2)={1 \over \left(1+\displaystyle{{\bf q}^2 \over \Lambda^2}\right)^2}
\,,
\ee
Pachucki \cite{pachucki1} obtains 0.021 meV (slightly larger in
\cite{pachucki2} or if one takes into account relativistic effects) 
using $\Lambda=848.5$ MeV. Within chiral perturbation theory, 
the Zemach correction to the muonic hydrogen Lamb shift in the limit $\Delta 
\rightarrow \infty$ reads
\be
\left.
\delta E_{\mu p}^{\rm Zemach}(n=2)\right|_{\Delta 
\rightarrow \infty}= -0.01008\,{\rm meV}
\,.
\ee
This is around 1/2 the standard predictions using the phenomenological
parameterization of the form factors. The inclusion of the effects due to 
the $\Delta$ increase this number by around a factor of two and bring it 
surprisingly close to the phenomenological estimates. 
\be
\delta E_{\mu p}^{\rm Zemach}(n=2)= -0.01922\,{\rm meV}
\,.
\ee
The corrections due to the Delta are large even if the series in $m_\pi/\Delta$
is convergent ($-0.01008-0.00561+\cdots$). It would be interesting to study 
whether something similar happens for the hyperfine splitting. 

Obviously, for the hydrogen we also obtain similar numbers to the 
phenomenological estimates if the $\Delta$ is included
\be
\delta E_{e p}^{\rm Zemach}(n=2)= -3.49652\,{\rm Hz}
\,.
\ee
The pure pionic correction reads
\be
\left.\delta E_{e p}^{\rm Zemach}(n=2)\right|_{\Delta 
\rightarrow \infty}= -1.83404\,{\rm Hz}
\,.
\ee

\medskip

The polarizability correction shown in sec. \ref{sec:matching} 
is only reliable for the case of hydrogen. 
It is proportional to the electric and magnetic proton polarizabilities. 
Therefore, the question is how good HBET can reproduce the experimental number 
for the polarizabilities. 
Actually, if only the pure "pionic" contribution is included a very good 
agreement is obtained. This agreement is deteriorated
if corrections due to the $\Delta$ are included (specially for $\beta_M$). In 
any case, the correction we obtain for the Lamb shift correction reads
\be
\left.\delta E_{e p}^{\rm pol.}\right|_{\Delta 
\rightarrow \infty}= -\frac{64.4841}{n^3}\,{\rm Hz}
\,,
\ee
and including the $\Delta$
\be
\delta E_{e p}^{\rm pol.}= -\frac{77.6037}{n^3}\,{\rm Hz}
\,.
\ee
We see that they are reasonable close to each other and with the value obtained 
using the experimental number for the proton polarizabilites \cite{KS}. Slightly 
larger numbers are obtained using  
the experimental data of the deep inelastic structure functions \cite{FM} or 
models \cite{Rosenfelder:1999px}. 

Finally, we would like to stress that in our determination there is no free parameter. 
The leading contribution is the one we have computed. 
This is due to the chiral enhancement. Our result diverges in the chiral and/or 
$1/N_c$ limit.

\medskip

For the vacuum polarization, the contribution due to the pion loop is 
numerically small (even if leading from the power counting point of view). 
One obtains $\Pi_{h,\pi}^{\prime} \simeq 1.0 \times \,10^{-3} \, {\rm GeV}^{-2}$. 
This should 
be compared with the experimental figure for the total hadronic contribution 
$\Pi_{h}^{\prime} \simeq 9.3 \times \, 10^{-3} \, {\rm GeV}^{-2}$ 
\cite{Jegerlehner:1996ab}. Actually, the main contribution comes from the 
$\rho$ even if it is suppressed in the power  counting ($\sim 6.1 \times \, 10^{-3}
\, {\rm GeV}^{-2}$). 
In any case we need $\Pi_{h}^{\prime}$ with $(\al/m_p^2)$ precision for 
which we can take the experimental number for the vacuum polarization (matching 
coefficient $d_{2,NR}$) and the correction to the Lamb shift reads
\be
\delta E_{ep}^{\rm vac.\;pol.} =-{3.39990 \over n^3}\,{\rm kHz}
\,,
\qquad 
\delta E_{\mu p}^{\rm vac.\;pol.}
=
-{0.09039 \over n^3}\,{\rm meV}
\,,
\ee
which for the hydrogen case reduces to the result obtained in Ref. \cite{Friar:1998wu}.

\section{Conclusions}

We have studied the lamb shift of hydrogen and muonic hydrogen within 
an effective field theory framework, connecting the physics at the hadronic 
scale (HBET) with the physics at the atomic scale (pNRQED). We have 
focused on the hadronic corrections and showed that a natural definition of the 
proton radius can be given as a matching coefficient of the effective theory. 

The leading hadronic corrections to the Lamb shift are a prediction of HBET 
(unknown couterterms are subleading). We have exactly computed them for the 
Zemach and vacuum polarizability terms, and for the polarizability term 
with logarithmic accuracy. 
For the Hydrogen atom the computed correction is beyond the present experimental 
accuracy. Nevertheless, the result is perfectly relevant for the muonic
hydrogen, for which it would be possible to get a model independent 
determination of the proton radius if 
the polarizability corrections were computed in HBET.

We have also done some partial checks of the (QED) hydrogen muonic computation. 
Nevertheless, it would be useful to perform a comprehensive numerical study 
for muonic hydrogen (since some pieces of the QED computation has only 
been performed by one group). Moreover, 
some further work remains to be done for the hadronic corrections: 
1) to obtain the complete result for the 
polarizability correction to the muonic hydrogen Lamb shift and 2) to perform the same 
analysis for SU(3) (including kaons). 

\medskip

{\bf Acknowledgments} \\ 
I would like to thank I. Scimemi and K. Pachucki for discussions.
This work is supported by MCyT and Feder
(Spain), FPA2001-3598, and by CIRIT (Catalonia), 2001SGR-00065.

\appendix

\Appendix{}
\label{app:formulas}

\bea
B_n&\equiv& \int_0^\infty \,dt
{t^{2-n} \over \sqrt{1-t^2}}\ln{\left[{1\over t}+\sqrt{{1 \over t^2}-1}\right]}
\\
\nn
&=&
-\frac{ {\sqrt{\pi }}\,{\Gamma}(\frac{3}{2} - \frac{n}{2})}{4\,
     {\Gamma}(2 - \frac{n}{2})}\,
       \left( {\rm HarmonicNumber}(\frac{1}{2} - \frac{n}{2}) - {\rm HarmonicNumber}(1 - \frac{n}{2}) \right)      
	 \\
	 \nn
	 &&
	 + 2^{1 - n}\,\pi \,{{\Gamma}( n-2) \over \Gamma^2(\frac{n}{2})}\,
   {{}_3F_2}( \frac{1}{2},\frac{ n-2}{2},\frac{ n-1}{2}; 
    \frac{n}{2},\frac{n}{2}; 1) 
   \\
   \nn
   &&
   - 
  \frac{2^{-2 + n}\,\pi }{{{\Gamma}(2 - \frac{n}{2})}^2}\,
  \Bigg( -3\,\left( n -2 \right) \,{\Gamma}(1 - n) + 2\,{\Gamma}(-n) 
  \\
  \nn
  &&
  \left.
  + 
       \frac{{\Gamma}(3 - n)}{\left(  n -3  \right) n}\,\left( -3 + 
	   \left( {n}-3 \right) \,{n}\,
	   \left( \pi \,\cot (n\,\pi ) - \log (2) - {\psi}(\frac{3}{2} - \frac{n}{2}) 
	   + {\psi}(n)\right)   
            \right)  \right)  
			\\
			\nn
			&&
			- 
  \frac{2}{n-3}\,
  \left.{\partial \over \partial x}\,{}_2F_{1}(1,\frac{-1 + n}{2};\frac{5 - n}{2}+x;-1)\right|_{x=0}
	\,,
\eea
where $\psi(n)=d\Gamma(n)/dn$ and $\Gamma(n)$ is the Euler $\Gamma$ function.
\be
H_n\equiv {n!(2n-1)!!\Gamma[-3/2] \over 2(2n)!!\Gamma[1/2+n]}  
\,.
\ee

\end{fmffile}


\begin{thebibliography}{99}

\bibitem{Taqqu} F. Kottmann et al., Hyperfine Interact. {\bf 138}, 55 (2001). 

\bibitem{pachucki1} K. Pachucki, Phys. Rev. {\bf A53}, 2092 (1996).

\bibitem{pachucki2} K. Pachucki, Phys. Rev. {\bf A60}, 3593 (1999).

\bibitem{pachucki4} A. Veitia and K. Pachucki, Phys. Rev. {\bf A69}, 042501 (2004).

\bibitem{CERN} J. \"Ayst\"o {\it et al.}, {\it Physics with low-energy 
muons at a neutrino factory complex}, hep-ph/0109217.

\bibitem{Karshenboim:1997zu}
S.~G.~Karshenboim,
Can.\ J.\ Phys.\  {\bf 77}, 241 (1999)
[arXiv:hep-ph/9712347].

\bibitem{Pachucki3} K. Pachucki, Phys. Rev. {\bf A52}, 1079 (1995). 

\bibitem{Mont} A. Pineda and J. Soto, Nucl. Phys. (Proc. Suppl.) {\bf 64}, 428
  (1998).  

\bibitem{lamb} A. Pineda and J. Soto, Phys. Lett. {\bf B420}, 391 (1998).
\bibitem{pos} A. Pineda and J. Soto, Phys. Rev. {\bf D59}, 016005 (1999). 
  
\bibitem{HF} A. Pineda, Phys. Rev. {\bf C67}, 025201 (2003); hep-ph/0308193. 

\bibitem{HBET} E. Jenkins and A.V. Manohar, Phys. Lett. {\bf B255}, 558
  (1991).  

\bibitem{Pineda:2000sz}
A.~Pineda and A.~Vairo,
Phys.\ Rev.\ D {\bf 63}, 054007 (2001)
[Erratum-ibid.\ D {\bf 64}, 039902 (2001)]
[arXiv:hep-ph/0009145].

\bibitem{Kinoshita:1979dy}
T.~Kinoshita and W.~B.~Lindquist,
Phys.\ Rev.\ D {\bf 27}, 853 (1983).

\bibitem{Borie:1982ax}
E.~Borie and G.~A.~Rinker,
Rev.\ Mod.\ Phys.\  {\bf 54}, 67 (1982).

\bibitem{Gasser:2003hk}
J.~Gasser, A.~Rusetsky and I.~Scimemi,
Eur.\ Phys.\ J.\ C {\bf 32}, 97 (2003)
[arXiv:hep-ph/0305260].

\bibitem{BFHM} V. Bernard, H.W. Fearing, T.R. Hemmert and U.-G.
Meissner, Nucl. Phys. {\bf A635}, 121 (1998); E-{\it ibid.}, {\bf
A642}, 563 (1998).  

\bibitem{Friar:1997tr}
J.~L.~Friar and G.~L.~Payne,
Phys.\ Rev.\ C {\bf 56}, 619 (1997)
[arXiv:nucl-th/9704032].

\bibitem{KS} I.B. Khriplovich and R.A. Sen'kov, Phys. Lett. {\bf A249}, 474 (1998). 

\bibitem{Bernard:1992qa}
V.~Bernard, N.~Kaiser, J.~Kambor and U.~G.~Meissner,
Nucl.\ Phys.\ B {\bf 388}, 315 (1992).

\bibitem{Hemmert:1999pz}
T.~R.~Hemmert, B.~R.~Holstein, G.~Knochlein and D.~Drechsel,
Phys.\ Rev.\ D {\bf 62}, 014013 (2000)
[arXiv:nucl-th/9910036].

\bibitem{BE} J. Bernabeu and T.E.O. Ericson, Z. Phys. {\bf A309}, 213
(1983). 

\bibitem{FM} R.N. Faustov and A.P. Martynenko, Phys. Atom. Nucl. {\bf
63}, 845 (2000), Yad. Fiz. {\bf 63}, 915 (2000).

\bibitem{Rosenfelder:1999px}
R.~Rosenfelder,
Phys.\ Lett.\ B {\bf 463}, 317 (1999)
[arXiv:hep-ph/9903352].

\bibitem{Jegerlehner:1996ab}
F.~Jegerlehner,
Nucl.\ Phys.\ Proc.\ Suppl.\  {\bf 51C}, 131 (1996)
[arXiv:hep-ph/9606484].

\bibitem{Friar:1998wu}
J.~L.~Friar, J.~Martorell and D.~W.~L.~Sprung,
Phys. Rev. A {\bf 59}, 4061 (1999)
[arXiv:nucl-th/9812053].

\end{thebibliography}
\end{document}